\documentclass[aip,reprint,nofootinbib]{revtex4-1}
\usepackage{graphicx}
\usepackage{dcolumn}
\usepackage{bm}

\begin{document}
\title{Transverse-Longitudinal Coupling by Space Charge in Cyclotrons}
\date{\today}
\author{C. Baumgarten}
\affiliation{Paul Scherrer Institute, Switzerland}
\email{christian.baumgarten@psi.ch}

\def\begeq{\begin{equation}}
\def\endeq{\end{equation}}
\def\begary{\begeq\begin{array}}
\def\endary{\end{array}\endeq}
\def\bmtx{\left(\begin{array}}
\def\emtx{\end{array}\right)}
\def\eps{\varepsilon}
\def\y{\gamma}
\def\w{\omega}
\def\W{\Omega}
\def\s{\sigma}

\def\Exp#1{\exp\left(#1\right)}
\def\Log#1{\ln\left(#1\right)}
\def\Sinh#1{\sinh\left(#1\right)}
\def\Sin#1{\sin\left(#1\right)}
\def\Tanh#1{\tanh\left(#1\right)}
\def\Tan#1{\tan\left(#1\right)}
\def\Cos#1{\cos\left(#1\right)}
\def\Cosh#1{\cosh\left(#1\right)}

\begin{abstract}
A method is presented that enables to compute the parameters of matched beams with space 
charge in cyclotrons with emphasis on the effect of the transverse-longitudinal coupling. 
Equations describing the transverse-longitudinal coupling and corresponding tune-shifts 
in first order are derived for the model of an azimuthally symmetric cyclotron. 
The eigenellipsoid of the beam is calculated and the transfer matrix
is transformed into block-diagonal form. The influence of the slope of the phase 
curve on the transverse-longitudinal coupling is accounted for.
The results are generalized and numerical procedures for the case of an azimuthally varying field 
cyclotron are presented. The algorithm is applied to the PSI Injector II and Ring
cyclotron and the results are compared to TRANSPORT.
\end{abstract}

\maketitle

\section*{Introduction}

There is continuous interest in the understanding of space charge effects in isochronous
cyclotrons~\cite{Reiser,Gordon5,Chasman,Adam1,Adam2,Bertrand,Pandit,SIR1,Bi}. The area is of special
relevance for the conceptual designs of cyclotrons as energy-efficient drivers for accelerator 
driven systems (ADS)~\cite{ADS0,ADS1}. 
Nevertheless there is no self-consistent first order theory of matched bunched beams 
with space charge in cyclotrons up to date known to the author. 
Bertrand and Ricaud showed that it is possible to set up and solve linearized equations 
of motion for a simple cyclotron model~\cite{Bertrand}, but they did not determine
the parameters of matched beams in cyclotrons and they did not generalize their results 
in order to find a numerical method to compute matched beams for sector-focused
cyclotrons. Information on the matched ellipsoid is of special interest for modern
high-performance codes like OPAL, that enable to simulate high intensity beams in the
space charge dominated regime~\cite{HPC1,HPC2}.

First we describe the usual simplified azimuthally symmetric cyclotron model 
(see for instance~\cite{Garren} or~\cite{Stammbach1}), but include the linearized space 
charge forces. The model neglects possible effects of rf-acceleration and is restricted
to the description of a coasting beam. 
A modified version of Teng's method to decouple longitudinal and horizontal motion will 
be presented and applied to the simplified analytical model~\cite{Teng,EdwardsTeng}.
Later we describe how this method can be applied in an iterative numerical
procedure to determine the parameters of matched beams with space charge in
sector-focused cyclotrons. Finally we present some results as computed for the 
cyclotrons of the PSI high intensity accelerator facility.

\section{The simplified cyclotron model}

The simplest relativistic cyclotron model is based on a magnetic field with
cylindrical and mid-plane symmetrie. The nominal orbital frequency of an ion
beam coasting in an isochronous cyclotron is 
\begeq
\w_o={q\over m\,\y}\,B={q\over m\,\y}\,B_0\,\y\,,
\endeq 
where $q$ is the charge and $m$ the rest mass of the ions. The axial magnetic field is $B=B(r)=B_0\,\y$. 
We define a cyclotron length unit $a={c\over\w_o}$ using the orbital frequency $\w_o$. 
The particle velocity is $v=\w_o\,r$ so that $\beta=v/c=r/a$. Hence the relativistic factor $\y$ 
can be written as
\begeq
\y={1\over\sqrt{1-(r/a)^2}}\,.
\endeq
The axial magnetic field in dependence of the radius $r$ is then given by
\begin{equation}
B(r)=B_0\,{1+\eps(r)\over\sqrt{1-{r^2\over a^2}}}\,,
\end{equation} 
where $\eps$ is a small distortion of the isochronism ($\eps\ll 1$).
The radial field increase compensates the relativistic mass increase
and the cyclotron is strictly isochronous for $\eps(r)=0$.
The field index $n={r\over B}\,{dB\over dr}$ is given by
\begeq
n={r\over B}\,{dB\over dr}\approx r\,{d\eps\over dr}+\y^2-1\,,
\endeq
so that
\begeq
1+n\approx\y^2+r\,{d\eps\over dr}\,.
\endeq
The phase shift per turn $\Delta\phi\equiv{d\phi\over dn}$ is (see for instance~\cite{Garren}:
\begary{rcl}
\Delta\phi&=&T\,{d\phi\over dt}=T\,\left({d\phi_{hf}\over dt}-N_h\,{d\theta\over dt}\right)\\
          &=&{2\,\pi\over\w_c}\,\left(\omega_{hf}-N_h\,\w_c\right)\\
          &=&{2\,\pi\over\w_c}\,\left(N_h\,\w_o-N_h\,\w_c\right)\\
          &=&2\,\pi\,N_h\,\left({\w_o\over\w_c}-1\right)\,,
\endary
where $\theta$ is the azimuthal angle, $\w_c=\dot\theta$ is the real orbital frequency and 
$\omega_{hf}=N_h\,\w_o$ the frequency of the accelerating high frequency 
system operated at the harmonic number $N_h$. The bending radius $r$ of a particle with
charge $q$ in a magnetic field $B$ is given by $r={p\over q\,B}$ and hence $\w_c$ can be written as
\begary{rcl}
\w_c&=&{v\over r}={m\,c\,\y\,\beta\over m\,\y\,r}={p\over m\,\y\,r}\\
        &=&{q\,B(r)\over m\,\y}={q\over m}\,{B_0\,\y\,(1+\eps)\over \y}=\w_o\,(1+\eps)\\
\endary
so that
\begary{rcl}
{\w_o\over\w_c}&=&{1\over 1+\eps}\approx 1-\eps\\
\eps&\approx&-{\Delta\phi\over 2\,\pi\,N_h}=-\left({\w_o\over\w_c}-1\right)\\
\endary
Hence ${d\eps\over dr}$ is proportional to the change of the phase shift per turn with radius.

The reference orbit corresponding to the kinetic energy $E=m\,c^2\,(\y-1)=E_0\,(\y-1)$ is a circle
with the radius 
\begeq
r(E)=a\,{\sqrt{\y^2-1}\over\y}\,.
\endeq
We will use $x=r-r(E)$ as the horizontal coordinate of a specific orbit relative to the reference orbit,
$y$ as the axial deviation from the median plane and $z$ as the longitudinal coordinate. 

\section{Some General Remarks}

From the assumption of midplane symmetrie it follows that the axial coordinate $y$ is not coupled
neither to the horizontal nor to the longitudinal coordinate. Therefore it is sufficient to
describe the coupling of horizontal and longitudinal motion. Since we assume azimuthal symmetrie, the 
matched beam ellipse must also be azimuthally symmetric, i.e. constant along the orbital length 
coordinate $s$. If the matched beam is constant, then the forces are constant - even if space
charge is taken into account.

The vector describing the (horizontal and longitudinal) displacement of an orbit at position $s$ 
relative to the reference orbit is ${\bf x}=(x,x',z,\delta)^T$ where $x'={dx\over ds}$ is the 
horizontal direction angle with the reference orbit and $\delta={p-p_0\over p_0}$ is the momentum 
deviation. We use the formalism of linear Hamiltonian systems as for instance described 
by R. Talman~\cite{Talman} including the method of A. Wolski to compute the matched beam 
matrix~\cite{Wolski}.
The equations of motion (to first order) can be written as
\begeq
{\bf x}'(s)={\bf F}\,{\bf x(s)}\,,
\endeq
where ${\bf F}$ is a constant $4\times 4$-matrix. The general solution is
\begeq
{\bf x}(s)=\exp{({\bf F}\,s)}\,{\bf x}(0)={\bf M}(s)\,{\bf x}(0)\,,
\endeq
where ${\bf M}(s)$ is the {\it transfer matrix}.
The matrix exponential ${\bf M}(s)$ is
\begeq
{\bf M}(s)=\exp{({\bf F}\,s)}={\bf 1}+{\bf F}\,s+{({\bf F}\,s)^2\over 2!}+{({\bf F}\,s)^3\over 3!}+\dots\,.
\endeq
Let $S$ be the skew-symmetric matrix with $S^2=-1$
\begeq
S=\bmtx{cccc}
0&1&0&0\\
-1&0&0&0\\
0&0&0&1\\
0&0&-1&0\\
\emtx\,,
\endeq
so that -- since ${\bf M}$ is sympletic -- the following relation holds:
\begeq
{\bf M}^T\,S\,{\bf M}=S\,.
\label{eq_symplectic}
\endeq
If there exists an invertible transformation matrix $E$ and a diagonal matrix $\lambda$
such that
\begeq
{\bf F}={\bf E}\,\lambda\,{\bf E}^{-1}\,,
\endeq
then its is easy to show that the following relation holds:
\begeq
{\bf M}={\bf E}\,\exp{(\lambda\,s)}\,{\bf E}^{-1}={\bf E}\,\Lambda\,{\bf E}^{-1}\,.
\label{eq_msplit}
\endeq
${\bf E}$ is the matrix of columnwise eigenvectors and $\lambda$ is the (diagonal) matrix of the corresponding
eigenvalues of ${\bf F}$. The (imaginary) eigenvalues of ${\bf F}$ are the betatron frequencies of the orbital 
motion. ${\bf F}$ and ${\bf M}$ have the same eigenvectors -- and the eigenvalues of ${\bf M}$ are the exponentials 
of the eigenvalues of ${\bf F}$. The eigenellipsoid $\sigma_E$ defined by 
\begeq
\s_E={\bf M}\,\s_E\,{\bf M}^T\,,
\label{eq_sigma_eigen_def}
\endeq 
can then be written as follows~\cite{Wolski}:
\begeq
\s_E=-{\bf E}\,{\cal D}\,{\bf E}^{-1}\,S\,,
\label{eq_sigma_eigen_split}
\endeq
where ${\cal D}$ is the diagonal matrix with the eigenvalues of $\s_E\,S$ -- which are the emittances (apart from factors $\pm i$).

\section{The Equations of Motion (EQOM)}

The Hamiltonian is given by
\begeq
H={x'^2\over 2}+{\delta^2\over 2\,\y^2}+{k_x-K_x\over 2}\,x^2-{\y^2\,K_z\over 2}\,l^2-h\,x\,\delta\,.
\endeq
where the canonical coordinates are $x$ and $l$ and the momenta are given by $x'$ and $\delta$.
The EQOM (in first order) are:
\begeq
{d\over ds}\,\bmtx{c}
x\\x'\\l\\\delta\\
\emtx=\bmtx{cccc}
0&1&0&0\\
-k_x+K_x&0&0&h\\
-h&0&0&{1\over\y^2}\\
0&0&K_z\,\y^2&0\\
\emtx\,\bmtx{c}x\\x'\\l\\\delta\\
\emtx\,,
\label{eq_eqom_sc}
\endeq
where $h={1\over r}$ is the inverse bending radius and $k_x=h^2\,(1+n)$ is the horizontally restoring force.
$K_x$ and $K_z$ represent the strength of the horizontal and longitudinal space charge forces~\cite{Hinterberger}:
\begary{rclp{10mm}rcl}
K_x&=&{K_3\,(1-f)\over (\sigma_x+\sigma_y)\,\sigma_x\,\sigma_z}&& K_3&=&{3\,q\,I\,\lambda\over 20\,\sqrt{5}\,\pi\,\eps_0\,m\,c^3\,\beta^2\,\y^3}\\
K_z&=&{K_3\,f\over \sigma_x\,\sigma_y\,\sigma_z}&&f&\approx&{\sqrt{\sigma_x\,\sigma_y}\over 3\,\y\,\sigma_z}\\
K_y&=&{K_3\,(1-f)\over (\sigma_x+\sigma_y)\,\sigma_y\,\sigma_z}&&&&\\
\label{eq_Kxyz}
\endary
The eigenvalues of ${\bf F}$ and ${\bf M}$ are:
\begary{rcl}
\lambda&=&\text{Diag}(i\,\W,-i\,\W,i\,\w,-i\,\w)\\
\Lambda&=&\text{Diag}(e^{i\,\W\,s},e^{-i\,\W\,s},e^{i\,\w\,s},e^{-i\,\w\,s})\\
a&\equiv&{k_x-K_x-K_z\over 2}\\
b&\equiv&K_z\,(K_x+h^2\,\y^2-k_x)\\
\W&=&\sqrt{a+\sqrt{a^2-b}}\\
\w&=&\sqrt{a-\sqrt{a^2-b}}\,.
\label{eq_freq}
\endary
Note that $b$ must be positive to give a real--valued frequency $\w$ and hence
a longitudinally focused orbit.
This is especially important, if we allow for a (small) field error $\eps(r)$.
In this case the change of the orbital frequency as given by ${d\eps\over dr}$ 
can play a significant role:
\begary{rcl}
k_x&=&h^2\,(1+n)=h^2\,(1+ r\,{d\eps\over dr}+\y^2-1)\\
   &=&h^2\,\left(\y^2+r\,{d\eps\over dr}\right)=h^2\,\y^2+h\,{d\eps\over dr}\\
b&=&K_z\,(K_x-h\,{d\eps\over dr})\\
\endary
If $K_x<h\,{d\eps\over dr}$, then the longitudinal focusing frequency $\w$ is imaginary
and the longitudinal beam size increases exponentially with $s$.
This in fact is a surprising feature of this type of coupling in combination
with isochronism: Longitudinal focusing requires a strong enough horizontally 
defocusing space charge force.
On the other hand, if ${d\eps\over dr}<0$, i.e. if the radial increase of the
magnetic field is below the isochronous field increase, then the longitudinal 
focusing is strengthened. 

The matrix of eigenvectors ${\bf E}$ of the force matrix ${\bf F}$ is
\begeq
{\bf E}=\bmtx{cccc}
1&1&1&1\\
i\,\W&-i\,\W&i\,\w&-i\,\w\\
i\,\W\,A&-i\,\W\,A&i\,\w\,B&-i\,\w\,B\\
K_z\,\y^2\,A&K_z\,\y^2\,A&K_z\,\y^2\,B&K_z\,\y^2\,B\\
\emtx\,.
\endeq
The inverse matrix ${\bf E}^{-1}$ is given by:
\begeq
{\bf E}^{-1}={1\over 2\,(A-B)}\bmtx{cccc}
-B&{i\,B\over\W}&-{i\over\W}&{1\over K_z\,\y^2}\\
-B&{-i\,B\over\W}&{i\over\W}&{1\over K_z\,\y^2}\\
A&-{i\,A\over\w}&{i\over\w}&-{1\over K_z\,\y^2}\\
A&{i\,A\over\w}&-{i\over\w}&-{1\over K_z\,\y^2}\\
\emtx\,,
\endeq
where 
\begary{rcl}
A&=&{h\over\W^2+K_z}\\ 
B&=&{h\over\w^2+K_z}\\
\label{eq_AB_def}
\endary
The transfer matrix ${\bf M}$ can be computed by using eq.~\ref{eq_msplit} and one obtains:
\begin{widetext}
\begeq
{\bf M}={1\over A-B}\,\bmtx{cccc}
A\,c-B\,C & {A\,\tilde s\over\omega}-{B\,S\over\Omega}&{S\over\Omega}-{\tilde s\over\omega}&A\,B\,(C-c)\\
\Omega\,B\,S-\omega\,A\,\tilde s  & A\,c-B\,C & C-c & A\,B\,(\omega\,\tilde s-\Omega\,S)\\
A\,B\,(\Omega\,S-\omega\,\tilde s) & A\,B\,(c-C) & A\,C-B\,c & A\,B\, (B\,\omega\,\tilde s-A\,\Omega\,S)\\
 c-C & {\tilde s\over\omega}-{S\over\Omega} &{S\over B\,\Omega}-{\tilde s\over A\,\omega} & A\,C-B\,c \\\emtx\,,
\endeq
\end{widetext}
where
\begary{rclp{10mm}rcl}
C&=&\cos{(\Omega\,s)} && S&=&\sin{(\Omega\,s)}\\
c&=&\cos{(\omega\,s)} && \tilde s&=&\sin{(\omega\,s)}\\
\endary

\section{The Eigenellipsoid}

The matrix ${\cal D}$ is explicitely given by ${\cal D}=\text{Diag}(i\,\eps_1,-i\,\eps_1,-i\,\eps_2,i\,\eps_2)$~\cite{Wolski}.
The matched eigenellipsoid can then be calculated using eq.~\ref{eq_sigma_eigen_split}:
\begin{widetext}
\begeq
\sigma_E={1\over B-A}\,\bmtx{cccc}
{B\,\eps_1\over\W}+{A\,\eps_2\over\w}&0&0&{\eps_1\over\W}+{\eps_2\over\w}\\
0   & B\,\eps_1\,\W+A\,\eps_2\,\w   & {\eps_1\,\W+\eps_2\,\w\over K_z\,\y^2}       & 0\\
0   & {\eps_1\,\W+\eps_2\,\w\over K_z\,\y^2} & {A\,\eps_1\,\W+B\,\eps_2\,\w\over K_z\,\y^2} & 0\\
{\eps_1\over\W}+{\eps_2\over\w}&0&0&{\eps_1\over B\,\W}+{\eps_2\over A\,\w}\\
\emtx
\endeq
\end{widetext}
The beam dimensions are given by the diagonal elements of the matrix representing the eigenellipsoid:
\begary{rcl}
\sigma_x^2&=&{1\over B-A}\left({B\,\eps_x\over\Omega}+{A\,\eps_z\over\omega}\right)\\
\sigma_z^2&=&{1\over B-A}\left({A\,\eps_x\,\Omega+B\,\eps_z\,\omega\over K_z\,\y^2}\right)\,,
\label{eq_sigxz}
\endary
where $\eps_1=\eps_x$ and $\eps_2=\eps_z$ are identified with the horizontal and longitudinal
emittance, respectively. The axial motion can be treated separately and one finds
\begeq
\sigma_y^2={\eps_y\over \sqrt{h^2\,\nu_y^2-K_y}}\,.
\label{eq_sigy}
\endeq
If one considers EQ.~\ref{eq_Kxyz} together with EQ.~\ref{eq_sigxz} and EQ.~\ref{eq_sigy}, then
it is obvious, that this result does not enable to start a straightforward calculation, since 
the beam sizes depend on the space charge forces and vice versa in an algebraically complicated 
way. But with the additional assumption of a spherical beam, it is possible to derive a 4th order
equation for the beam size as will be shown in Sec.~\ref{sec_spherical}.

\section{Decoupling Longitudinal and Transverse Motion}

If sectors are considered, then the azimuthal symmetrie is broken and hence the force terms in the 
EQOM and consequently the beam ellipsoid depend on the position $s$.
The vertical beam dynamics can still be treated separately, but is has to be taken into account that
the periodic change of the vertical beam size influences the space charge factors $K_x$ and $K_z$. 
The design orbit usually has to be computed numerically as described by Gordon~\cite{Gordon}.
If this has been done, it is possible to compute the transfer matrix for known starting conditions, 
since the equations of motion are known and can be integrated. The problem is to find the 
correct beam dimensions for a given beam current and given emittances such that the beam is matched,
i.e. such that EQ.~\ref{eq_sigma_eigen_def} is fulfilled. 

Teng and Edwards described a parametrization for coupled motion in two and more dimensions
that allows to find the decoupling matrix~\cite{Teng,EdwardsTeng}. They called their method 
{\it symplectic rotation}. We will give a brief summary of the method and apply it to the 
problem of decoupling longitudinal and transverse motion. 

Given two $2\times 2$ symplectic matrices ${\cal A}$ and ${\cal B}$ that form a block-diagonal 
(i.e. decoupled) transfer matrix $T_0$:
\begary{rcl}
T_0&=&\bmtx{cc}{\cal A}&0\\0&{\cal B}\emtx\\
{\cal A}(\mu_1)&=&\cos{(\mu_1)}\,{\bf 1}+\sin{(\mu_1)}\,\bmtx{cc}\alpha_1&\beta_1\\-\y_1&-\alpha_1\emtx\\
{\cal B}(\mu_2)&=&\cos{(\mu_2)}\,{\bf 1}+\sin{(\mu_2)}\,\bmtx{cc}\alpha_2&\beta_2\\-\y_2&-\alpha_2\emtx\,,
\label{eq_T0_def}
\endary
where $\alpha_i$, $\beta_i$ and $\y_i$ are the familiar twiss-parameters. 
Then a general symplectic transfer matrix ${\bf M}$ of coupled motion can be written as
\begeq
{\bf M}=\bmtx{cc}M&n\\m&N\emtx={\bf R}\,T_0\,{\bf R}^{-1}\,.
\endeq
where $M$, $N$, $m$ and $n$ are $2\times 2$ matrices and ${\bf R}$ is the symplectic rotation matrix.
Teng suggested to write ${\bf R}$ in the form
\begary{rcl}
{\bf R}&=&\bmtx{cc}
I\,\cos{(\phi)}&D^{-1}\,\sin{(\phi)}\\
-D\,\sin{(\phi)}&I\,\cos{(\phi)}\emtx\\
{\bf R}^{-1}&=&\bmtx{cc}
I\,\cos{(\phi)}&-D^{-1}\,\sin{(\phi)}\\
D\,\sin{(\phi)}&I\,\cos{(\phi)}\emtx\,.
\endary
where $D$ is a symplectic $2\times 2$ transfer matrix that describes the structure of the coupling.
Then one obtains~\cite{Teng,EdwardsTeng}:
\begary{rcl}
M&=&{\cal A}\,\cos^2{(\phi)}+D^{-1}\,{\cal B}\,D\,\sin^2{(\phi)}\\
N&=&{\cal B}\,\cos^2{(\phi)}+D\,{\cal A}\,D^{-1}\,\sin^2{(\phi)}\\
m&=&-(D\,{\cal A}-{\cal B}\,D)\,\sin{(\phi)}\,\cos{(\phi)}\\
n&=&-({\cal A}\,D^{-1}-D^{-1}\,{\cal B})\,\sin{(\phi)}\,\cos{(\phi)}\\
\endary
The rotation angle $\phi$ can be computed using
\begary{rcl}
\frac{1}{2}\,Tr(M-N)&=&\cos{(2\,\phi)}\,(\cos{(\mu_1)}-\cos{(\mu_2)})\\
2\,Det(m)+Tr(n\,m)&=&\sin^2{(2\,\phi)}\,(\cos{(\mu_1)}-\cos{(\mu_2)})^2\,,
\endary

If we apply this method to the transfer matrix computed according to EQ.~\ref{eq_msplit}, we obtain:
\begary{rcl}
\frac{1}{2}\,Tr(M-N)&=&{B+A\over B-A}\,(\cos{(\Omega s)}-\cos{(\omega s)})\\
2\,Det(m)+Tr(n\,m)&=&-{4\,A\,B\over (B-A)^2}\,(\cos{(\Omega s)}-\cos{(\omega s)})^2\,,
\endary
Comparison yields:
\begary{rcl}
\cos{(2\,\phi)}&=&{B+A\over B-A}\\
\sin^2{(2\,\phi)}&=&-{4\,A\,B\over (B-A)^2}\\
\endary
From EQ.~\ref{eq_AB_def} we find that $A>0$ and $B>0$ and $B>A$ since $\Omega^2>\omega^2$, so 
that the method fails in the case under study. The method of symplectic rotation is not generally 
applicable and has to be extended. A workaround solution was found by replacing trigonometic
by hyperbolic functions\footnote{Formally this extension is a {\it rotation about an imaginary angle}, 
similar to a Lorentz boost in Minkowski space.}.
\begary{rcl}
{\bf R}&=&\bmtx{cc}
I\,\cosh{(\psi)}&D^{-1}\,\sinh{(\psi)}\\
D\,\sinh{(\psi)}&I\,\cosh{(\psi)}\emtx\\
{\bf R}^{-1}&=&\bmtx{cc}
I\,\cosh{(\psi)}&-D^{-1}\,\sinh{(\psi)}\\
-D\,\sinh{(\psi)}&I\,\cosh{(\psi)}\emtx\,,
\endary
with $Det(D)=-1$, i.e. $D$ is {\it not symplectic} -- but ${\bf R}$ is symplectic.  
\begin{table}
\begin{tabular}{l|l|l}
                     & Symplectic Rotation  &  ``Symplectic Boost'' \\\hline
$Tr(M-N)/2$          & $\cos{(2\,\phi)}\,\Delta$      & $\cosh{(2\,\psi)}\,\Delta$\\
$D_t$                & $\sin{(2\,\phi)}^2\,\Delta^2$  & $-\sinh{(2\,\psi)}^2\,\Delta^2$\\
$D$                  & $-{m+S_2\,n^T\,S_2^T\over \Delta\,\sin{(2\,\phi)}}$ & ${m+S_2\,n^T\,S_2^T\over\Delta\,\sinh{(2\,\psi)}}$\\
${\cal A}$           & $M-D^{-1}\,m\,\tan{(\phi)}$    & $M-D^{-1}\,m\,\tanh{(\psi)}$\\
${\cal B}$           & $N+D\,n\,\tan{(\phi)}$         &  $N-D\,n\,\tanh{(\psi)}$\\
\end{tabular}
\caption{
Comparison of the method of symplectic rotation with the symplectic ``Lorentz boost''.
The difference $\Delta$ is defined by $\Delta\equiv \cos{(\mu_1)}-\cos{(\mu_2)}$. The matrix
$S_2$ is defined by $S_2=\bmtx{cc}0&1\\-1&0\emtx$ and $D_t$ is defined by $D_t=2\,Det(m)+Tr(n\,m)$.
\label{tab_rot_lt}
}
\end{table}
Tab.~\ref{tab_rot_lt} compares the formulas of the symplectic rotation and of symplectic ``Lorentz boost''.
Applying this method to the case under study then gives:
\begary{lclp{5mm}lcl}
\sinh{(2\,\psi)}&=&{2\,\sqrt{A\,B}\over B-A}&&\cosh{(2\,\psi)}&=&{B+A\over B-A}\\
\tanh{(2\,\psi)}&=&{2\,\sqrt{A\,B}\over B+A}&&\sinh{(\psi)}&=&\sqrt{A\over B-A}\\
\cosh{(\psi)}&=&\sqrt{B\over B-A}&&\tanh{(\psi)}&=&\sqrt{A\over B}\,,
\endary
which can be solved. The matrices  ${\cal A}$ and  ${\cal B}$ are then given by:
\begary{rcl}
{\cal A}&=&\bmtx{cc}\cos{(\Omega s)}&\sin{(\Omega s)}/\Omega\\-\Omega\,\sin{(\Omega s)}&\cos{(\Omega s)}\emtx\\
{\cal B}&=&\bmtx{cc}\cos{(\omega s)}&-A\,B\,\omega\,\sin{(\omega s)}\\\sin{(\omega s)}/(A\,B\,\omega)&\cos{(\omega s)}\emtx\\
D&=&\bmtx{cc}0&\sqrt{A\,B}\\1/\sqrt{A\,B}&0\emtx\\
\endary
The signs in ${\cal B}$ suggest a negative $\beta_2$ - but this can be compensated by using either a negative 
emittance or a negative frequency $\omega$, since the transfer matrix is invariant against a change of the sign 
of $\omega$, while $\sigma_E$ is not. 

The transformation matrices are explicitely given by:
\begary{rcl}
{\bf R}&=&{\sqrt{B\over B-A}}\,
\bmtx{cccc}
1&0&0&A\\
0&1&1/B&0\\
0&A&1&0\\
1/B&0&0&1\\
\emtx\\
&&\\
{\bf R}^{-1}&=&{\sqrt{B\over B-A}}\,
\bmtx{cccc}
1&0&0&-A\\
0&1&-1/B&0\\
0&-A&1&0\\
-1/B&0&0&1\\
\emtx\,.
\endary
Note: ${\bf R}$ and ${\bf R}^{-1}$ are symplectic. To make the method complete, we give the formula to split
$2\times 2$ matrices of the form of ${\cal A}$ or ${\cal B}$:
\begary{rcl}
c&=&\cos{\mu}\\
s&=&\sin{\mu}\\
\bmtx{cc}
c&s/K\\
-K\,s&c\\
\emtx&=&R_2\,
\bmtx{cc}
e^{-i\,\mu}&0\\
0&e^{i\,\mu}\\
\emtx\,R_2^{-1}\\
R_2&=&\bmtx{cc}
i/K&-i/K\\
1 & 1\\
\emtx\\
R_2^{-1}&=&
\bmtx{cc}
-i\,K/2&1/2\\
i\,K/2&1/2\\
\emtx\\
\label{eq_diag2}
\endary
The matched beam ellipsoid can be computed for given emittances according to EQ.~\ref{eq_sigma_eigen_split}
as soon as the diagonali\-zation of the transfer matrix ${\bf M}$ is known.

The fact that we have to extend the decoupling method of Teng and Edwards raises the question, whether
the description of decoupling is now complete or if there are other cases that require further modifications or
extensions. The answer to this question requires a proper two-dimensional extension of the Courant-Snyder theory
and a complete survey of all possible symplectic transformations as presented in an accompanying paper~\cite{rdm_paper}.

\section{The Iteration Process and Examples}

We have shown for the symmetric analytical example that a modified version of the method of Teng and Edwards 
enables to bring the transfer matrix in block-diagonal form according to EQ.~\ref{eq_T0_def}. The 
diagonali\-zation of the block-diagonal matrices is given by EQ.~\ref{eq_diag2} and hence the matrix of
eigenvectors $E$ and the matched beam ellipsoid can be constructed. 

In order to take advantage of this method for the case of sectored cyclotrons, the method has to be
applied iteratively. The goal is to compute the properties of a matched beam. 
Assumed that the 
beam emittances and the current are given as boundary conditions, one proceeds as follows:
\begin{enumerate}
\item Provide an initial guess of the beam dimensions $\sigma_x(s)$, $\sigma_y(s)$ and $\sigma_z(s)$.
\item Compute the driving terms of the space charge forces $K_x(s)$, $K_y(s)$ and $K_z(s)$.
\item Compute the one-turn transfer matrix ${\bf M}(s)$ for all azimuthal angles.
\item Compute the eigenvectors of ${\bf M}(s)$ by diagonali\-zation.
\item Use the beam emittances to obtain the eigenellipsoid $\sigma_E$ according to EQ.~\ref{eq_sigma_eigen_split}. 
\item Take the beam sizes from $\sigma_E$ and go back to step 2, 
      if beam sizes changed significantly compared to the previous iteration.  
\end{enumerate}
Optionally one can start the iteration with a reduced beam current and increase it either during the iteration
process or use the result of the reduced beam current as an initial guess for higher currents.
The (speed of) convergence strongly depends on the assumptions about beam current and emittance.
In the following examples, the process converged typically after less than 20 iterations.

\subsection{Example: PSI Injector II}

The PSI high intensity proton accelerator facility~\cite{MikePierre,Mike2008} (HIPA) consists of a 
Cockcroft-Walton pre-accelerator, a four-sector injector cyclotron (Injector II, $72\,\mathrm{MeV}$) 
and the eight sector Ring cyclotron ($590\,\mathrm{MeV}$). In routine operation the beam current is
$2.2\,\mathrm{mA}$.

\begin{figure}
\includegraphics[width=8.5cm]{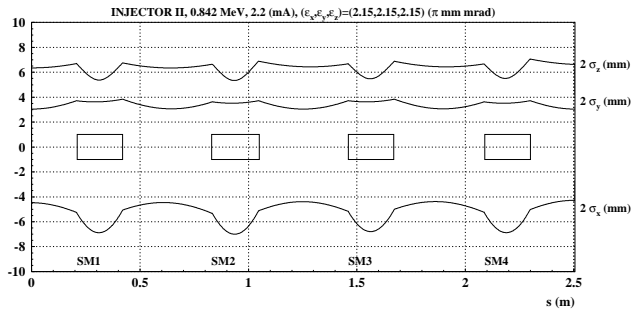}
\caption{
Transport (re-)computation of a matched beam in Injector II for a circulating beam of
$2.2\,\mathrm{mA}$ at low energy. The boxes indicate the position of the sector magnets SM1..SM4.
The horizontal beam width is plotted as a negative value. The longitudinal
beam size is plotted as $2\,\sigma_z$.
\label{fig_inj2trans}
}
\end{figure}

The Injector II cyclotron can be modelled reasonably well by a hard edge approximation of the sector magnets.
This has the advantage that the computed matched beam parameters can be compared to TRANSPORT~\cite{transport1,transport2,transport3}. We have computed the matched beam parameters for Injector II. The results have then be
used as input parameters for TRANSPORT (with space charge). The results are shown in Fig.~\ref{fig_inj2trans}.

\subsection{Example: PSI Ring Cyclotron}

The PSI ring cyclotron is a separated sector isochronous cyclotron with 8 sectors. But since the
field of the magnets does not fall off as sharp as it does in Injector II, a hard edge approximation
does not work equally well. Furthermore the sectors are much more spiralled. For the sake of precision
we used the measured field map $B(r,\theta)$ to compute the equilibrium orbits (EO) on a radial grid~\cite{eo_paper}. 
\begin{figure}
\includegraphics[width=8.5cm]{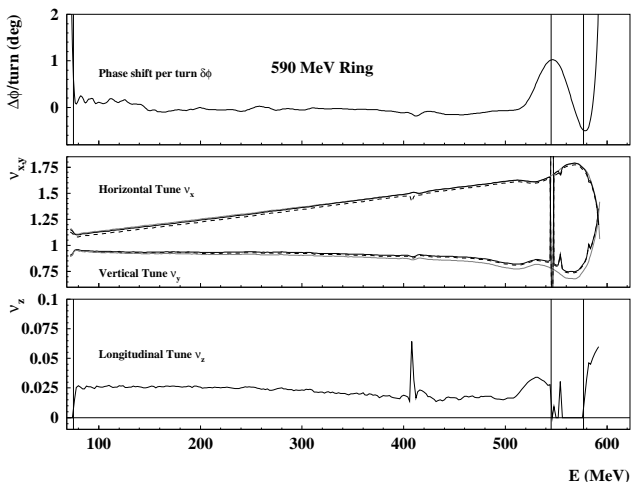}
\caption{
Top: Phase shift per turn for the ring cyclotron computed from the measured magnetic field map. 
Middle: Horizontal and vertical betatron-tunes of the Ring-cyclotron as computed by EO-code (gray solid lines),
the transfer matrix method over the EO (black solid lines) and with space charge (dashed lines).
Bottom: Longitudinal betatron tune as induced by transverse--longitudinal coupling.
The region with a negative slope of the phase shift is indicated by vertical lines.
Without further field trimming, the longitudinal focusing is suppressed by the 
term $-h\,{d\eps\over dr}$ and the longitudinal beam size may increase.
The calculation has been performed for a beam current of $2.2\,\mathrm{mA}$ and
$(\eps_x,\eps_y,\eps_z)=(1.5,2.5,0.5)\,\mathrm{\pi mm mrad}$.
\label{fig_ring_eo}
}
\end{figure}
The radius $R_{eo}(\theta)$ of the equilibrium orbit (EO) is then given as a function of the angle
and of the starting radius $R_{eo}(\theta_0)$. Besides the pure geometry of the orbit one also 
obtains a precise value for $\eps=1-{\omega_o\over\omega}$ for each EO~\cite{Gordon}, which then
allows to compute ${d\eps\over dr}$.
In order to obtain the force matrix according to EQ.~\ref{eq_eqom_sc}, 
the following additional calculations have been done:
\begin{enumerate}
\item Inverse bending radius $h={1\over\rho}={B(r,\theta)\over B_0\,p}$
\item Local field index $$n={\rho\over B}\,{dB\over dx}={B_0\over B^2}\,\left({dB\over dr}\,p_\theta-{dB\over r\,d\theta}\,p_r\right)\,.$$
\item Path length element $\Delta s=\sqrt{r'^2+r^2}\,\Delta\theta$.
\item Horizontal (vertical) focusing $k_x={1+n\over\rho^2}$ ($k_y={-n\over\rho^2}$).
\item $h(E,\theta)$, $n(E,\theta)$, $\y(E)$ and the space charge forces $K_x$ , $K_y$ , $K_z$ are then
used to compose the force matrix ${\bf F}$.
\item The exponential series allows to compute the transfer matrix for a (short) interval 
$[\theta\dots\theta+\Delta\theta]$:
\begeq
{\bf M}=\exp{(F\,\Delta s)}=1+F\,\Delta s+{F^2\over 2!}\,\Delta s^2+\dots
\endeq
(Typically $3\dots 4$ terms are sufficient.)
\item Multiplication of all matrices yields the one-turn transfer matrix ${\bf M}$.
\item Diagonalize ${\bf M}$ and compute eigenellipse $\sigma_E$.
\item The average beam sizes are obtained from the eigenellipsoid and used to recompute 
the force coefficients.
\item The convergence is then checked and either the next iteration starts with step 5) or is stopped.
\end{enumerate}
Fig.~\ref{fig_ring_eo} shows some results of the matched beam with space charge for the PSI ring cyclotron,
the phase shift per turn vs. energy in the upper, the transversal tunes in the second and the axial tune
in the last graph. The region where the {\it slope} of the phase shift is negative has been marked by
vertical lines. In this region the longitudinal tune due to space charge forces apparently becomes 
imaginary.

\section{Spherical Symmetry}
\label{sec_spherical}

The first item in the described algorithm is to provide an initial guess of the
beam dimensions. In the following we show how to obtain a possible choice of the 
initial guess for ``nearly'' spherical beams. If the horizontal-longitudinal 
coupling is strong enough, the beam will usually be approximately circular in this 
plane~\cite{Stammbach2}. The axial size depends mainly on the vertical emittance 
and tune.

From EQ.~\ref{eq_freq} and EQ.~\ref{eq_AB_def} one quickly derives:
\begary{rcl}
\W^2\,\w^2&=&K_z\,(K_x+h^2\,\y^2-k_x)\\
\W^2+\w^2&=&k_x-K_x-K_z\\
\W^2-\w^2&=&\sqrt{(k_x-K_x+K_z)^2-4\,h^2\,\y^2\,K_z}\\
{1\over B-A}&=&{\y^2\,h\,K_z\over\sqrt{(k_x-K_x+K_z)^2-4\,h^2\,\y^2\,K_z}}\\
\endary
With the assumption of an isochronous cyclotron ($k_x=h^2\,\y^2$) and spherical 
symmetrie, i.e.
\begary{rcl}
\sigma&=&\sigma_x=\sigma_y=\sigma_z\,\y\\
\eps&=&\eps_x=\eps_y=\eps_z\\
\rightarrow&&K=K_x=K_y=K_z={K_3\,\y\over 3\,\sigma^3}\\
\endary
one finds
\begary{rcl}
\W^2\,\w^2&=&K^2\\
\W^2+\w^2&=&k_x-2\,K\\
\W^2-\w^2&=&\sqrt{k_x^2-4\,K\,k_x}\\
{1\over B-A}&=&{K\over h\,\sqrt{1-4\,K/k_x}}\\
\endary
The beam size $\sigma$ then yields using EQ.~\ref{eq_sigxz}:
\begary{rcl}
\sigma^4&=&\eps^2\,{(B\,\w+A\,\W)^2\over (\W\,\w)^2\,(B-A)^2}\\
        &=&{4\,\eps^2\over k_x-4\,K}\,,
\endary
so that one obtains
\begeq
\sigma^4-{4\,K_3\,r^2\over 3\,\y}\,\sigma-{4\,\eps^2\,r^2\over\y^2}=0\,.
\label{eq_4th_order}
\endeq
If one defines $\sigma_0=\sqrt{2\,r\,\eps\over\y}$, then EQ.~\ref{eq_4th_order}
can be written as
\begeq
x^4-\alpha\,x-1=0\,,
\label{eq_4poly}
\endeq
where 
\begary{rcl}
x&=&{\sigma\over\sigma_0}\\
\alpha&=&{4\,K_3\,r^2\over 3\,\y\,\sigma_0^3}={K_3\,\sqrt{2\,\y\,r}\over 3\,\eps^{3/2}}>0\,.
\endary
A polynom of 4th order has 4 solutions, namely the solutions of EQ.~\ref{eq_4poly} are 
\begeq
x={1\over 2\,6^{1/3}}\,\left(\sqrt{Y}\pm\,\sqrt{\pm\,{12\,\alpha\over\sqrt{Y}}-Y}\right)\,,
\label{eq_pol4sol}
\endeq
where the abbreviations used are defined by
\begary{rcl}
Y&=&2^{1/3}\,Z-{8\,3^{1/3}\over Z}\ge 0\\
Z&=&(9\,\alpha^2+\sqrt{3}\,\sqrt{256+27\,\alpha^2})^{1/3} > 768^{1/6}\,.
\endary
The minus sign in the square root of EQ.~\ref{eq_pol4sol} belongs to a pair of complex 
conjugate solutions and may not be used. Starting from low values of $\alpha$, $Y$ 
monotonically increases starting from zero so that we have to take the plus in front 
of the square root as well in order to obtain positive solutions also for small values 
of $\alpha$, so that
\begeq
x={1\over 2\,6^{1/3}}\,\left(\sqrt{Y}+\sqrt{{12\,\alpha\over\sqrt{Y}}-Y}\right)\,.
\endeq
Analyzing EQ.~\ref{eq_4poly} one finds that
\begary{rcl}
\lim\limits_{\alpha\to 0}\,x(\alpha)&=&1\\
\lim\limits_{\alpha\to\infty}\,x(\alpha)&=&\alpha^{1/3}\,,
\endary
so that the function $x(\alpha)=(1+\alpha)^{1/3}$ is a reasonable approximation.
The maximal relative deviation of about $-0.035$ appears at $\alpha\approx 3/2$.
Therefore we suggest to use the approximations
\begeq
\sigma\approx\left\{\begin{array}{lcl}
\sigma_0\,(1+{\alpha\over 4}-{\alpha^2\over 32})&\mathrm{for}&0\le\alpha\le 5/2\\
\sigma_0\,(1+\alpha)^{1/3}&\mathrm{for}&5/2\le\alpha\\
\end{array}\right.\,.
\label{eq_sig_a}
\endeq
Using the normalized emittance $\bar\eps=\beta\,\y\,\eps$, the wavelength $\lambda={c\over\nu_{rf}}={2\,\pi\,c\over\omega_o\,N_h}$, 
the cyclotron radius $a={c\over\omega_0}={r\over\beta}$ and the relation $\eps_0\,c^2={1\over\mu_0}$, the values of 
$K_3$, $\alpha$ and $\sigma_0$ can be written as follows:
\begary{rcl}
K_3&=&{3\,q\,I\,\lambda\over 20\,\sqrt{5}\,\pi\,\eps_0\,m\,c^3\,\beta^2\,\y^3}\\
   &=&{3\,q\,\mu_0\,I\,a\over 10\,\sqrt{5}\,m\,c\,\beta^2\,\y^3\,N_h}\\
\alpha&=&{q\,\mu_0\,I\,a\,\sqrt{2\,\y\,r}\over 10\,\sqrt{5}\,m\,c\,\beta^2\,\y^3\,\eps^{3/2}\,N_h}\\
      &=&{q\,\mu_0\,I\over 5\,\sqrt{10}\,m\,c\,\y\,N_h}\,\left({a\over\bar\eps}\right)^{3/2}\\
\sigma_0&=&\sqrt{2\,r\,\eps\over\y}={\sqrt{2\,a\,\beta\,\y\,\eps}\over\y}={\sqrt{2\,a\,\bar\eps}\over\y}\,.
\label{eq_alp_def}
\endary
EQ.~\ref{eq_sig_a} and \ref{eq_alp_def} describe the {\it matched} beam sizes for a spherical beam in a perfectly 
isochronous cyclotron. The {\it real} beam sizes will usually differ from these values since the spherical
symmetrie requires that the vertical tune roughly equals the horizontal tune, while in most cyclotrons the 
vertical tune is below the horizontal tune. Nevertheless the equations derived for the special case of a
spherical beam can be used as starting conditions for the described iterative matching procedure.

\section{Summary}

A method has been developed that allows to compute the parameters of a matched beam 
with space charge in cyclotrons in linear approximation for given beam current and
known emittances. As an example, a matched beam in the PSI INJECTOR II cyclotron has been
computed. The result has been used as input to TRANSPORT and it has been shown that the
beam is matched. 

Furthermore it has been shown that the longitudinal focusing as induced by the space charge forces 
depends on the isochronism of the cyclotron. In case of the PSI Ring cyclotron, the negative slope 
of the phase shift per turn may reduce or even destroy the longitudinal focusing effect. 
The results are important for the design of high intensity cyclotrons. If beam emittance and current are
chosen accordingly, then the space charge induced longitudinal focusing allows to replace 
flat-top cavities by accelerating cavities and thus increase the energy gain per turn and 
the turn separation significantly. This is important to minimize beam losses and activation 
of components. The flat-top cavities of the PSI Injector II cyclotron are already operating
as accelerating cavities and a complete replacement is planned~\cite{Stammbach2,Bopp}.

\section{Acknowledgements}

Mathematica\textsuperscript{\textregistered} $5.2$ has been used for symbolic calculations.
Software has been written in ``C'' and been compiled with the GNU\textsuperscript{\copyright}-C++
compiler 3.4.6 on Scientific Linux.

\end{document}